\documentclass[preprint]{JHEP3} 

\JHEPspecialurl{http://jhep.sissa.it/JOURNAL/JHEP3.tar.gz}
\usepackage{graphicx}
\usepackage{amsmath,amssymb}
\usepackage{cite}
\usepackage{multirow}

\newcommand{\eps}{\epsilon}

\title{Gluon fusion contribution 
to  $W^+W^- + \mathrm{jet}$ production}

\author{Tom Melia \\Rudolf Peierls Centre for Theoretical Physics, 1
  Keble Road, University of
  Oxford, UK\\
  Email: \email{t.melia1@physics.ox.ac.uk} }

\author{Kirill~Melnikov \\Department of Physics and Astronomy, Johns
  Hopkins
  University, Baltimore, MD 21218, USA\\
  Email: \email{melnikov@pha.jhu.edu} }

\author{Raoul R\"ontsch \\Rudolf Peierls Centre for Theoretical
  Physics, 1 Keble Road, University of
  Oxford, UK\\
  Email: \email{r.rontsch1@physics.ox.ac.uk} } 

\author{Markus Schulze \\High Energy Physics Division, Argonne National Laboratory, Argonne, IL 60439, USA \\
 Email: \email{markus.schulze@anl.gov}
}

\author{Giulia Zanderighi \\Rudolf Peierls Centre for Theoretical
  Physics, 1 Keble Road, University of
  Oxford, UK\\
  Email: \email{g.zanderighi1@physics.ox.ac.uk} }

\received{\today} 		
\accepted{\today}		

\preprint{OUTP-12-07P} 

\abstract{We describe the computation of the $gg \to W^+W^-g$ process
  that contributes to the production of two $W$-bosons and a jet at
  the CERN Large Hadron Collider (LHC).  While formally of
  next-to-next-to-leading order (NNLO) in QCD, this process can be
  evaluated separately from the bulk of NNLO QCD corrections because
  it is finite and gauge-invariant.  It is also enhanced by the large
  gluon flux and by selection cuts employed in the Higgs boson
  searches in the decay channel $ H \to W^+W^-$, as was first pointed
  out by Binoth {\it et al.} in the context of $gg \to W^+W^-$
  production.  For cuts employed by the ATLAS collaboration, we find
  that the gluon fusion contribution to $pp \to W^+W^-j$ enhances the
  background by about ten percent and can lead to moderate 
  distortions of kinematic distributions which are instrumental for the
  ongoing Higgs boson searches at the LHC. We also release a public code to compute the NLO QCD
  corrections to this process, in the form of an add-on to the package {\tt MCFM}.}


\begin{document} 
\section{Introduction}

At the time of writing, the Large Hadron Collider (LHC) is running at a center-of-mass energy $\sqrt{s}=8$ TeV, having completed a very
successful $7$ TeV run at the end of 2011. Collected data are
consistent with the Standard Model Higgs boson with a mass of
$124-126$ GeV \cite{ATLAS:2012ae,Chatrchyan:2012tx}. As more data are
collected in 2012, the evidence for the new particle in this mass
range will, hopefully, become stronger, opening up the way for a detailed
exploration of its properties.

Currently, the $m_H = 125~{\rm GeV}$ Higgs boson signal has the
largest significance in $\gamma \gamma$ and $Z Z^*$ final states.
However, eventually, the signal should also appear in the $W^+W^-$ final
state thanks to the process $gg \to H \to W^+W^-$. In fact,
understanding the properties of the new particle will require, at a
minimum, the measurement of its branching fractions to as many final states as possible; therefore, measuring the
branching fraction for $H \to W^+W^-$ is important.  Higgs searches in
the $W^+W^-$ mode are performed by binning final states according to
the number of jets produced. This is primarily done for a detailed
identification of backgrounds and for designing cuts that maximize the
signal-to-background ratio for each of the jet bins.  For zero- and
one-jet bins, the dominant backgrounds are $pp \to W^+W^-$ and $pp \to
W^+W^-+{\rm jet}$, respectively, while for the two-jet bin $pp \to t
\bar t $ becomes dominant.

QCD radiative corrections are known to change both signal and
background rates by significant, jet-bin-dependent amounts. Higgs
production rates are known through next-to-next-to-leading order
(NNLO) QCD in the zero-jet
bin~\cite{Harlander:2002wh,Anastasiou:2002yz,Ravindran:2003um} and
through NLO QCD in the
one-~\cite{deFlorian:1999zd,Ravindran:2002dc,Glosser:2002gm} and
two-jet bins~\cite{Campbell:2006xx,Campbell:2010cz}.  The production
of a $W$-pair without reconstructed jets is known through NLO QCD
\cite{Ohnemus:1991kk,Frixione:1993yp,Dixon:1998py,Dixon:1999di,Campbell:1999ah}
and has been implemented in publicly available programs such as {\tt
  MCFM} \cite{Campbell:2011bn}, {\tt MC@NLO}~\cite{Frixione:2002ik},
and {\tt POWHEG} \cite{Melia:2011tj}. The production of a $W$-pair in
association with one jet was calculated through NLO QCD in
Refs. \cite{Campbell:2007ev,Dittmaier:2007th,Dittmaier:2009un}.
The NLO QCD corrections to the production of a $W$-pair in association
with two jets were computed in
Refs. \cite{Melia:2010bm,Melia:2011dw,Greiner:2012im}.

Little is known about the main background processes beyond next-to-leading
order.  One exception is the study of the gluon-gluon fusion that
gives rise to a pair of $W$-bosons through a fermion loop
\cite{Kao:1990tt,Binoth:2005ua,Binoth:2006mf}. Although this
contribution appears at one-loop, it cannot contribute to
next-to-leading order cross-sections because there is no tree process
with two gluons in the initial state. Therefore, the $gg \to W^+W^-$
sub-process is finite and a gauge-invariant part of the NNLO
contribution to $W^+W^-$ production that can be computed
independently.  
Thanks to the low production threshold of the two $W$-bosons relative
to the LHC center-of-mass energy, the cross-section arising from the
$gg \to W^+W^-$ contribution is enhanced by the large gluon flux. 
To some extent, this may compensate the suppression by an additional
power of $\alpha_s$ and lead to the numerical importance of the gluon
fusion sub-process.  These NNLO corrections have been studied for the
production of a $W$-pair with no jets in
Refs.~\cite{Kao:1990tt,Binoth:2005ua,Binoth:2006mf}.  It was found
that the effect of these corrections is highly cut-dependent: for
inclusive cuts they enhance the cross-section by $4-5 \%$, which is
comparable to the scale uncertainty at next-to-leading order, but the relative
importance of the gluon fusion corrections increases dramatically if
cuts designed to enhance the $gg \to H \to W^+W^-$ signal are applied.
In fact, for the cuts of Refs. \cite{Binoth:2005ua,Binoth:2006mf}, $gg
\to W^+W^-$ becomes the {\it dominant} radiative correction to $pp \to
W^+W^-$, overshooting the NLO QCD correction to this process by a
factor of around seven.
The gluon induced WW production is implemented in the generators {\tt
  MCFM}~\cite{Campbell:2011bn} and {\tt gg2WW}~\cite{Binoth:2006mf};
the latter can also be interfaced to parton shower programs.

\begin{figure}[t]
\begin{center}
\includegraphics[scale=0.7]{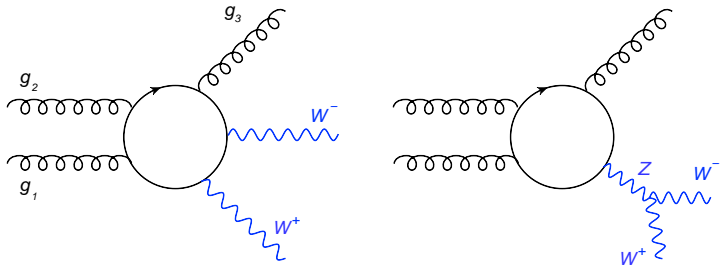}
\caption{Sample Feynman diagrams for $gg \to W^+W^-g $ through a
  fermion loop.} \label{fig:fermloop}
\end{center}
\end{figure}

Since, as we pointed out earlier, the Higgs searches distinguish
processes with different jet multiplicities and since the final state
with a $W$-pair and a jet is an important contributor to current Higgs
searches,
it is a relevant
question if a similar enhancement of the gluon fusion contribution
occurs there as well.  The goal of this paper is to study this
question.  To this end, we compute the gluon-induced NNLO QCD
corrections to $W^+W^- + $ jet production in proton collisions and
explore its numerical significance in dependence of the applied cuts.

To perform this calculation, we use a generalized unitarity
implementation~\cite{Ellis:2007br,Giele:2008ve,Ellis:2008ir,Ellis:2011cr}
of the Ossola-Pittau-Papadopoulos (OPP) procedure~\cite{Ossola:2006us}
for the reduction of tensor integrals (for a review see
Ref.~\cite{Ellis:2011cr}).  These on-shell techniques have had
fantastic success over the past few years, leading to a large number
of NLO QCD computations of very complex
processes~\cite{Maestre:2012vp}.
Recent developments in automating these and similar techniques for
one-loop computations also seem very
promising~\cite{Hirschi:2011pa,Bevilacqua:2011xh,Cullen:2012pj,Cascioli:2011va}.

The remainder of this paper is organized as follows. In Section
\ref{sec:techdet}, we provide technical details of the calculation,
pointing out a few subtleties that arise when dealing with a process
that involves several colorless external bosons and virtual fermions.
In Section \ref{sec:results}, we present results for typical cuts that
are used to observe weak gauge boson pair production and for a
different set of cuts designed to increase the significance of the
Higgs boson signal.  We conclude in Section \ref{sec:concl}. In the
Appendix, we provide numerical results for one-loop primitive
amplitudes for further comparison.  Finally, we note that as a
by-product of this work we release a code that allows for the NLO QCD
description of the $pp \to W^+W^-+$jet process  \footnote{Note that the public code does not include the NLO QCD corrections coming from fermion loops -- these contributions are several orders of magnitude smaller than the total cross-sections.}.  The code is in the form
of a patch to the program {\tt MCFM}~\cite{Campbell:2011bn}, and can be
obtained from \href{http://www-thphys.physics.ox.ac.uk/people/TomMelia/tommelia.html}{http://www-thphys.physics.ox.ac.uk/people/TomMelia/tommelia.html}, or from the {\tt MCFM} website \href{http://mcfm.fnal.gov/}{http://mcfm.fnal.gov/}.

\begin{figure}[t]
\begin{center}
\includegraphics[scale=0.5]{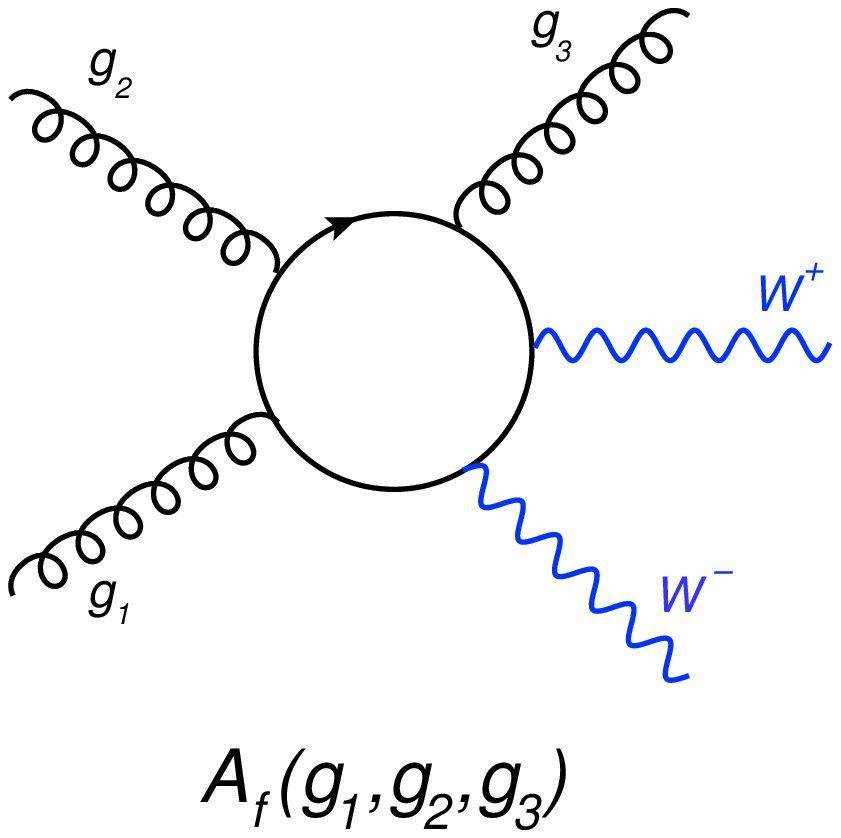}\hspace*{1cm}
\includegraphics[scale=0.5]{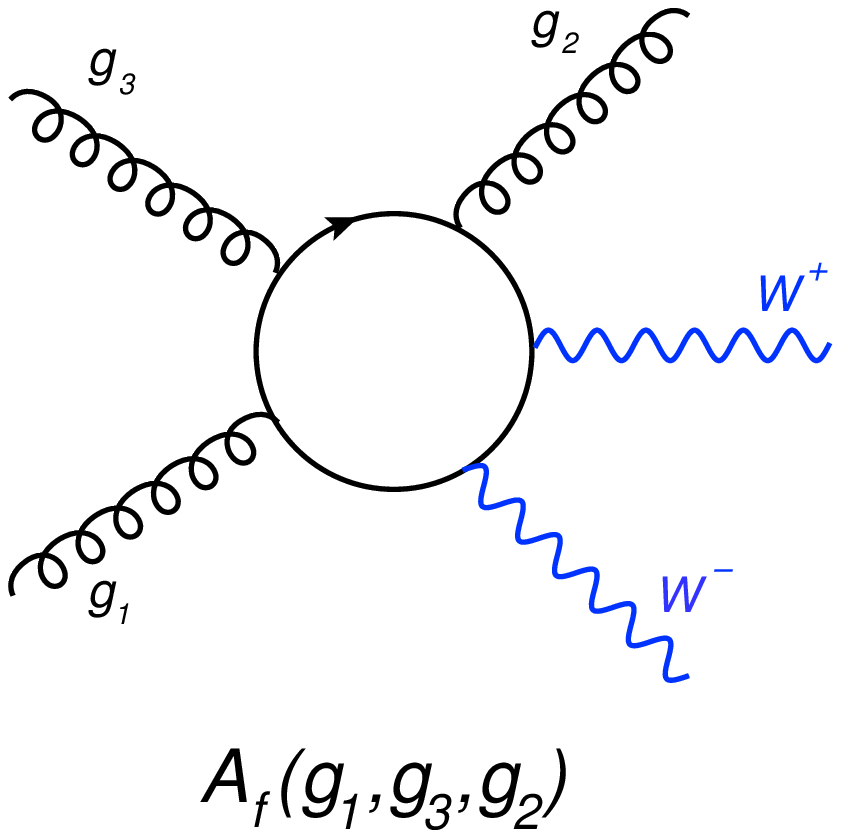}\\
\caption{Primitive amplitudes $A_f(g_1,g_2,g_3)$ and
  $A_f(g_1,g_3,g_2)$. One ordering of the $W$-bosons is shown. However, we need to consider every insertion of the electroweak bosons
  relative to the ordered gluons.} \label{fig:partamp}
\end{center}
\end{figure}

\section{Technical Details} 
\label{sec:techdet} 
We consider the gluon-gluon fusion contribution to the production of a
$W$-boson pair and a jet. At lowest order in QCD, this process occurs
through a fermion loop and involves three gluons, see
Fig.~\ref{fig:fermloop}.  This one-loop amplitude provides a finite,
gauge-invariant contribution to the NNLO correction.  We consider
leptonic decays of $W$ bosons and write the one-loop amplitude for the
process $0 \to g g g W^+(\to \nu_e e^+) W^-(\to \mu^-
\bar{\nu}_{\mu})$ as
\begin{equation} \label{partamp}
\begin{split}
 \mathcal{A}_f^{1}(g_1,g_2,g_3; &\nu_e, e^+,\mu^-,\bar{\nu}_{\mu})  = g_s^3 \left ( \frac{g_w}{\sqrt{2}} \right )^4 \times \\
&  \Bigl( \mathrm{Tr}\bigl(T^{a_1} T^{a_2} T^{a_3} \bigr) A_f(g_1,g_2,g_3) + \mathrm{Tr}\bigl(T^{a_1} T^{a_3} T^{a_2} \bigr) A_f(g_1,g_3,g_2) \Bigr).
\end{split}
\end{equation}
Here $g_w$ and $g_s$ are the weak and strong couplings respectively,
and the $A_f$ are color-ordered primitive amplitudes with all 
possible insertions of the two W bosons relative to the ordered
gluons, see Fig.~\ref{fig:partamp}.  The normalization of the $SU(3)$ color group generators is $\mathrm{Tr}(T^a T^b) = \delta^{ab}$. We take top and bottom quarks as
massive, while the other four quarks are taken to be massless. We use
a unit CKM matrix, so that each of the primitive amplitudes in
Eq.~\eqref{partamp} contains two massless and one massive fermion
loop,
\begin{equation} \label{0m}
 A_f = 2 A_{f,0} +  A_{f,m}.
\end{equation} 
It is also necessary to consider the production of $WW$ through an
intermediate $Z$-boson or a photon. Thus we can write the primitive
amplitudes in Eq.~\eqref{0m} as
\begin{equation} \label {WWZgamma}
\begin{split}
 A_{f,0} &= A_{f,ud}^{[WW]} + \sum_{q\in\{u,d\}} \Bigl( C_{V,q}^{[Z]} A_{f,q}^{[Z_V]} + C_{A,q}^{[Z]} A_{f,q}^{[Z_A]} + C_q^{\gamma} A_{f,q}^{[\gamma]} \Bigr)\,,\\
 A_{f,m} &= A_{f,tb}^{[WW]}  + \sum_{q\in\{t,b\}} \Bigl( C_{V,q}^{[Z]} A_{f,q}^{[Z_V]} + C_{A,q}^{[Z]} A_{f,q}^{[Z_A]} + C_q^{\gamma} A_{f,q}^{[\gamma]} \Bigr)\,.
\end{split}
\end{equation}
The first term refers to the amplitudes obtained by attaching the
$W$-bosons directly to the fermion loop. The remaining terms refer to
amplitudes obtained for one flavor of quark circulating in the loop,
with the $W$-bosons attached through the vector and axial parts of a $Z$-boson
or a photon, respectively. The couplings are
$C_{V,q}^{[Z]} = T^3_q - 2Q_q \sin^2 \theta_w$, $C_{A,q}^{[Z]} =
-T^3_q$, and $C_{q}^{[\gamma]} = 2Q_q \sin^2 \theta_w$, where $Q_q$ is
the electromagnetic charge of the quark in the loop, $\theta_w$ is the
weak mixing angle, and $T^3_{u,c,t} = 1/2, T^3_{d,s,b} = -1/2$.
For the first term in Eq.~\eqref{WWZgamma}, it is necessary to
consider every insertion of the $W$-pair relative to the ordered
gluons. There are twelve such insertions, allowing for the ordering
$W^+W^-$ and $W^-W^+$.  Similarly, for the remaining terms, we need to
account for three orderings of an intermediate $Z$-boson or photon
relative to the gluons.
We account for the decays of $W$-bosons $W (p_l + p_{\bar{l}}) \to
l(p_l) \bar{l} (p_{\bar{l}})$ by constructing polarization vectors
from lepton spinors.  We write
\begin{equation}
 \eps_{W}^{\mu}(p_l,p_{\bar{l}}) = 
\frac{-\mathrm{i} \bar{u}(p_l)\gamma^{\mu} 
(1-\gamma_5)
 v(p_{\bar{l}})
}{2(s_{l \bar l} - m_W^2 + \mathrm{i} \Gamma_W m_W)}, 
\end{equation}
where $s_{l \bar{l}} = (p_l + p_{\bar{l}})^2$ is the invariant mass of
the two leptons.  The definitions of the polarization vectors of the
intermediate $Z$-boson and the photon include a large number of terms
that accommodate both single and double resonant amplitudes for the
production of two $W$-bosons.  All amplitudes were checked against 
an independent home-made OPP-based program that can be used 
to compute Feynman  diagrams directly.
Furthermore, 
the term $A_{f,ud}^{[WW]}$  was checked against the 
publicly available program {\tt GoSam} \cite{Hirschi:2011pa} for 
a few phase space points. 

\begin{figure}[t]
\begin{center}
\includegraphics[scale=0.5]{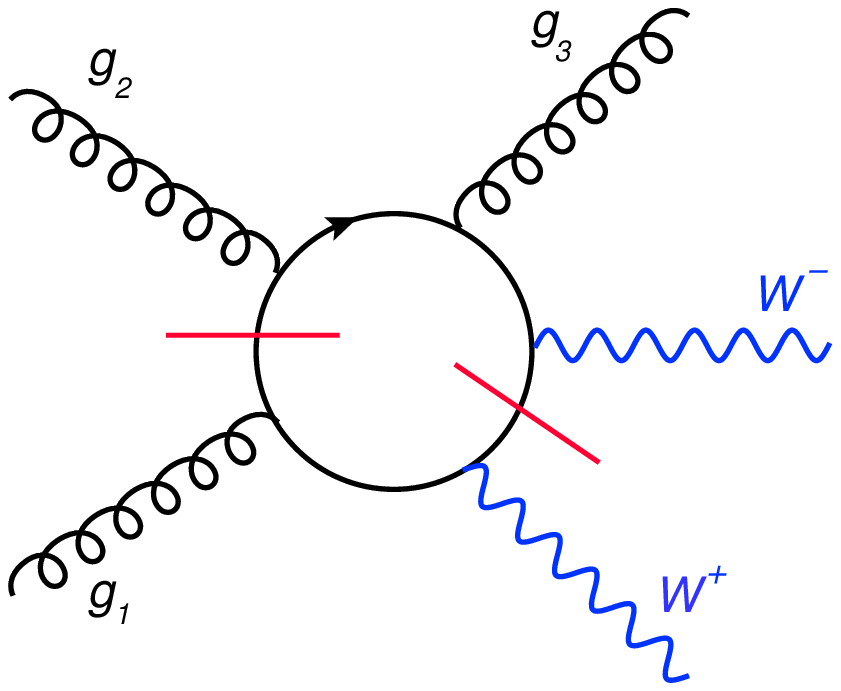}\hspace*{1cm}
\includegraphics[scale=0.5]{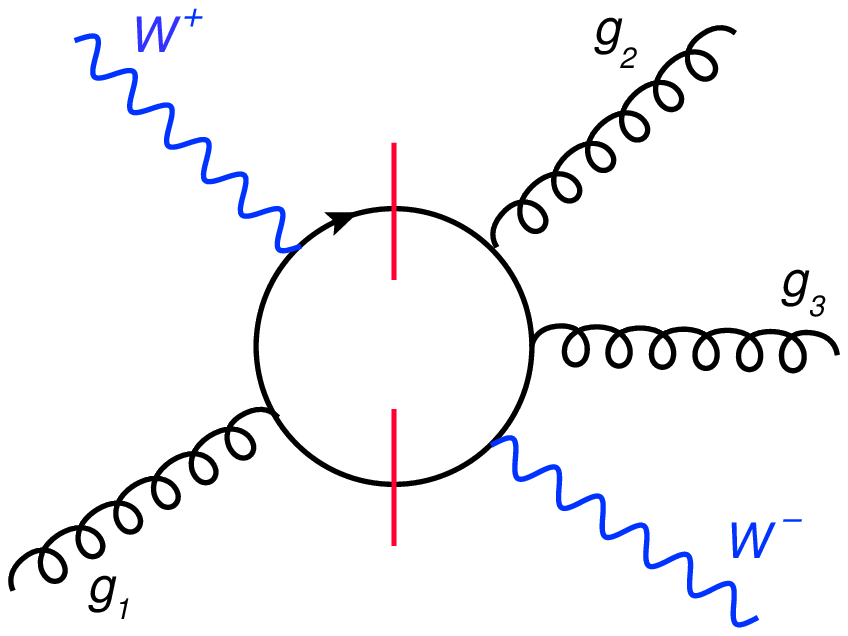}
\caption{Two double cuts on a primitive amplitude with different
  orderings of the $W$-bosons, leading to a potential source of double
  counting.} \label{fig:doubcut}
\end{center}
\end{figure}

We compute tree level amplitudes using Berends-Giele recursion
relations \cite{Berends:1989}.  There is a further technical aspect to
be discussed, related to the use of these recursion relations with a
cyclic amplitude. To illustrate this point, we consider two orderings
of the $W$-bosons in the primitive amplitude $A_f(g_1,g_2,g_3)$, with
a different double cut on each, as shown in
Fig.~\ref{fig:doubcut}. Both of these cuts split the one-loop
amplitude into two tree-level helicity amplitudes, and we focus on the
one involving $g_1$ and $W^+$. When a Berends-Giele current includes
an electroweak boson, all insertions of that electroweak boson
relative to the gluons are considered. In particular, this means that
the tree-level amplitude arising from either cut shown will include
the $W^+$ both before and after the gluon $g_1$, reading
clockwise. Since the amplitudes are cyclic, we would, in effect, be
double counting. Because the cuts are on different propagators, it is
difficult to identify automatically which cuts will end up duplicating
one another in this manner.
Instead, we observe that the cause of this problem is the cyclic
nature of the amplitudes. However, looking at Eq.~\eqref{partamp}, it
is clear that we can require gluon $g_1$ to be in a fixed position for
both primitive amplitudes, with appropriate ordering of the other two
gluons and all insertions of the $W$-bosons. We modify our
Berends-Giele currents accordingly, so that the positions of the
electroweak bosons are fixed relative to $g_1$, but still allowing all
insertions relative to the other two gluons.  Referring again to
Fig.~\ref{fig:doubcut}, the left amplitude will have the $W^+$-boson
only before the gluon (again, reading clockwise), while the right
amplitude will have the $W^+$-boson only after the gluon, and the double
counting is avoided.

\section{Results} 
\label{sec:results}
In this Section we present the results for the process $pp \to W^+W^-
+ {\rm jet}$ at the LHC, including the gluon fusion contribution $gg
\to W^+W^-g$.  We consider two center-of-mass energies, $\sqrt{s}=8$
GeV and $\sqrt{s}=14$ TeV and let the $W$-bosons decay into a pair of
leptons of definite flavor, $W^+W^- \to \nu_e e^+ \mu^-
\bar{\nu}_{\mu}$. We note that the cross-section inclusive in lepton
flavors $l^+ l^- = \{ e^+e^-, e^+ \mu^-, \mu^+ e^-, \mu^+\mu^-\}$ is a
factor of four larger than the results presented below.
We take the $W$-boson mass and width to be $m_W = 80.399$ GeV and
$\Gamma_W = 2.085$ GeV, while the $Z$-boson mass and width are taken
to be $m_Z = 91.1876$ GeV and $\Gamma_Z = 2.4952$ GeV. We use a top
quark mass $m_t = 172.9$ GeV and a bottom quark mass $m_b = 4.19$
GeV and take all other quarks to be massless. The weak couplings are defined
through the Fermi constant $G_F = 1.166364 \times 10^{-5}$ GeV$^{-2}$
with $g_w^2 = 8G_F m_W^2/\sqrt{2}$, and the weak mixing angle is given
by $\sin^2 \theta_w = 1-m_W^2/m_Z^2$. Jets are defined through the
anti-$k_\perp$ algorithm~\cite{Cacciari:2008gp} with $R=0.4$, as
implemented in FastJet~\cite{Cacciari:2011ma}. We use MSTW08 parton
distribution set~\cite{Martin:2009iq} and we employ leading,
next-to-leading and next-to-next-to-leading order parton distribution
functions (pdf) to compute the relevant contributions to the
cross-sections.  We stress that, for the purposes of this discussion,
the gluon fusion process $gg \to W^+W^-g$ is a NNLO contribution and
therefore it is computed with NNLO pdfs\footnote{We note that the use
  of NNLO pdfs for the computation of the gluon fusion contribution
  leads to a smaller result (by roughly $30\%$) compared to
  that obtained using LO pdfs, since the NNLO
  gluon flux is smaller than the LO flux.  We nevertheless use NNLO
  pdfs since this is what would be done once the NNLO description of
  the $W^+W^-$ production becomes available.}.

\begin{table} 
\begin{center}

Standard Cuts 

\vspace{0.3cm}

\begin{tabular}{|c|c|c|c|c|c|c|}
\hline
& &$\sigma_{\mathrm{LO}}$ & $\sigma_{\mathrm{NLO}}^{\mathrm{incl}}$ & $\sigma_{\mathrm{NLO}}^{\mathrm{excl}}$ & $\delta \sigma_{\mathrm{NNLO}}$ & $ \delta \sigma_{\mathrm{NNLO}} / \sigma_{\mathrm{NLO}}^{\mathrm{incl}}$ \\
\hline
\multirow{2}{*}{8 TeV}  & $WW$  & $141.0(1)^{+2.8}_{-4.0} $   &
$232.0(4)^{-5.8}_{+7.5}$  &$143.8(2)^{+4.2}_{-4.1}$     &
$8.1(1)^{-1.7}_{+2.2}$ & $3.5 \% $\\ 
                        & $WW$j  & $87.8(1)^{-10.9}_{+13.5}$    &
                        $111.3(2)^{-5.5}_{+4.9}$
                        &$66.6(2)^{+4.4}_{-9.0}$     &
                        $3.4(1)_{+1.6}^{-1.0}$ & $3.1 \% $\\
\hline
\multirow{2}{*}{14 TeV} & $WW$  & $259.6(2)^{+14.2}_{-17.2}$   &
$448.3(5)^{-7.4}_{+11.6}$ &$242.0(3)^{+9.2}_{-8.6}$    &
$23.6(1)^{-4.1}_{+5.2}$ & $5.3 \% $\\
                        & $WW$j& $203.4(1)^{-19.9}_{+22.9}$   &
                        $254.5(4)^{-10.2}_{+9.0}$
                        &$127.6(4)^{+14.8}_{-24.1}$      &  $11.8(4)
                        ^{-3.2}_{+4.7}$ & $4.6 \% $\\
\hline
\end{tabular}
\end{center}
\caption{Cross-sections for $pp \to W^+(\nu_e e^+)W^-(\mu^- \bar \nu_\mu ) + n$ jets, $n=0,1$ at the $\sqrt{s} = 8$ TeV and $\sqrt{s}=14$ TeV LHC, 
  using the standard cuts described in the text. All cross-section values are in femtobarns. The central values are computed 
  with  factorization and renormalization scales $\mu = 2m_W$, with the 
  statistical error shown  
  in parentheses. The upper and lower values for cross-sections are obtained with the choice 
  of scale 
  $\mu = m_W$ (subscript) and $\mu = 4m_W$ (superscript). The last column shows the relative size 
  of the NNLO contribution and the \textit{inclusive} NLO
  cross-section for the central scale choice.}
\label{tab:xsincl}
\end{table}

We first consider cuts that are typical for processes with $W$-bosons
and jets, and refer to them as ``standard cuts''. We require that {\it
  i}) both leptons have transverse momenta $ p_{T,l} > 20$ GeV and
pseudorapidity $| \eta_l | < 2.5$, {\it ii}) the missing transverse
momentum satisfies $p_{T,\mathrm{miss}} > 30$ GeV and {\it iii}) jets
have transverse momenta $p_{T,j} > 20$ GeV and pseudorapidity
$|\eta_j| < 3.2$.  In Table \ref{tab:xsincl}, we show the LO and NLO
cross-sections for $pp \to W^+W^- + $ jet, as well as the gluon fusion
(NNLO) contribution, using these cuts for two LHC energies.  Also
shown are the sizes of the NNLO corrections relative to the inclusive
NLO cross-sections. We set the factorization and renormalization
scales equal to one another, $\mu_F = \mu_R = \mu$, and show the scale
variation $m_W \leq \mu \leq 4m_W$ with a central scale choice $\mu =
2m_W$.  We also show the cross-sections for $pp \to W^+W^-$ without
jets in Table~\ref{tab:xsincl}, for comparison with the earlier
calculation of Ref.~\cite{Binoth:2005ua}.  Although the setup here is
slightly different,
in particular cuts on missing transverse energy, parton distribution
functions and some input parameters, our
results are similar to the ones reported in Ref.~\cite{Binoth:2005ua}
for LO, NLO and the gluon fusion contributions to $pp \to W^+W^-$
process.  It follows from Table~\ref{tab:xsincl} that the gluon fusion
corrections are modest; they change the NLO cross-sections by $3-5\%$
which is to be compared with ${\cal O}(65-70\%)$ inclusive NLO QCD
corrections in the case of $pp \to W^+W^-$ and ${\cal O}(15-25\%)$ in the case
of $pp \to W^+W^-+{\rm jet}$.  The relative importance of the gluon
fusion contribution grows with the energy which is to be expected because
of the rapid growth of the gluon flux at small values of $x \sim
m_{W^+W^-}/\sqrt{s}$.  For both processes, $pp \to W^+W^-$ and $pp \to
W^+W^-+j$, the gluon fusion corrections are comparable to the scale
uncertainty of the NLO cross-section at $\sqrt{s} = 8~{\rm TeV}$ and
slightly exceed it at $\sqrt{s} = 14~{\rm TeV}$.

Next, we consider cuts similar to those used by the ATLAS
collaboration in their Higgs search in the $W^+W^-$ channel
\cite{ATLAS:2012} (``Higgs search cuts''). To this end, we require
that {\it i}) the hardest lepton has $p_{T,l,\mathrm{max}} > 25 $ GeV,
the softest lepton has $p_{T,l, \mathrm{min}} > 15~{\rm GeV}$, {\it
  ii}) the lepton rapidities are $|\eta_l| < 2.5$~\footnote{In the
  measurements of ATLAS~\cite{ATLAS:2012} the rapidity region $1.37 <
  |\eta| < 1.52$ is excluded for electrons. For simplicity we use
  instead full acceptance both for electrons and muons. We also ignore issues related to specifics of the detector. Our
treatment of the lepton separation is also slightly simpler than the one in
Ref.~\cite{ATLAS:2012}.}; {\it iii})
the relative azimuthal angle between the two leptons is small $\Delta
\phi_{ll} < 1.8$, {\it iv}) the two leptons are separated from each
other by $\Delta R_{ll} = \sqrt{ \Delta \phi_{ll}^2 +
  \Delta\eta_{ll}^2} > 0.3$, 
{\it v}) the invariant mass of the charged lepton system must satisfy $10$ GeV
$< m_{ll} < 50$ GeV, {\it vi}) the missing relative transverse
energy satisfies $E_{T,\mathrm{rel}}^{\mathrm{miss}} > 25$ GeV.  We
note that the missing relative transverse energy is defined as
\begin{equation}
 E_{T,\mathrm{rel}}^{\mathrm{miss}} = | {\bf p}_{T}^{\mathrm{miss}} | \sin \Delta \phi_{\mathrm{min}},
\end{equation}
where $\Delta \phi_{\mathrm{min}} = \min
\bigl(\Delta\phi_{\mathrm{miss}}, \pi/2 \bigr)$ and
$\Delta\phi_{\mathrm{miss}}$ is the azimuthal angle between the
missing transverse momentum vector ${\bf p}_{T}^{\mathrm{miss}}$ and the
momentum of the nearest charged lepton or jet with $p_T >25$ GeV. Jets are identified using the anti-$k_\perp$
algorithm with $R=0.4$ and $p_{T,j} > 25~{\rm GeV}$.  All jets must
have pseudorapidities in the region $| \eta_j | < 4.5$.
For the $WW$ process we require additionally that the charged lepton
system has a transverse momentum $p_{T,ll} > 30$ GeV. 

The cross-sections obtained using these cuts are shown in Table
\ref{tab:xsatlas}.  While the importance of NLO QCD corrections is
considerably reduced with these cuts, the relative importance of gluon
fusion corrections increases by a factor of two, compared to the
standard cuts. The change to the {\it exclusive} NLO cross-sections
for $pp \to W^+W^-$ and $pp \to W^+W^-+j$ because of the gluon fusion
contribution is about $6\%$ for $\sqrt{s} = 8~{\rm TeV}$ and is close
to $10\%$ for $\sqrt{s} = 14~{\rm TeV}$. Therefore, for the Higgs search 
cuts, gluon fusion corrections are larger than the scale variations of
 the exclusive NLO QCD cross-sections for the process $W^+W^-$, and comparable to 
the scale variations  of
 the exclusive NLO QCD cross-sections for the process $W^+W^-+1$~jet. 

 Furthermore, using the Higgs search cuts described above and a Higgs
 mass of $m_H = 125$ GeV, the \textit{signal} $H\to WW$ cross-section
 (in the $e^+ \mu^-$ channel considered here) at NLO QCD is
 approximately $5$ fb ($12$ fb) at $8$ TeV ($14$ TeV) in the $0$-jet
 channel and $2$ fb ($5$ fb) at $8$ TeV ($14$ TeV) in the $1$-jet
 channel. The gluon fusion corrections to the $WW$ background alone
 therefore amount to almost half of the signal cross-sections.

\begin{figure}[t]
\begin{center}
\includegraphics[scale=0.6]{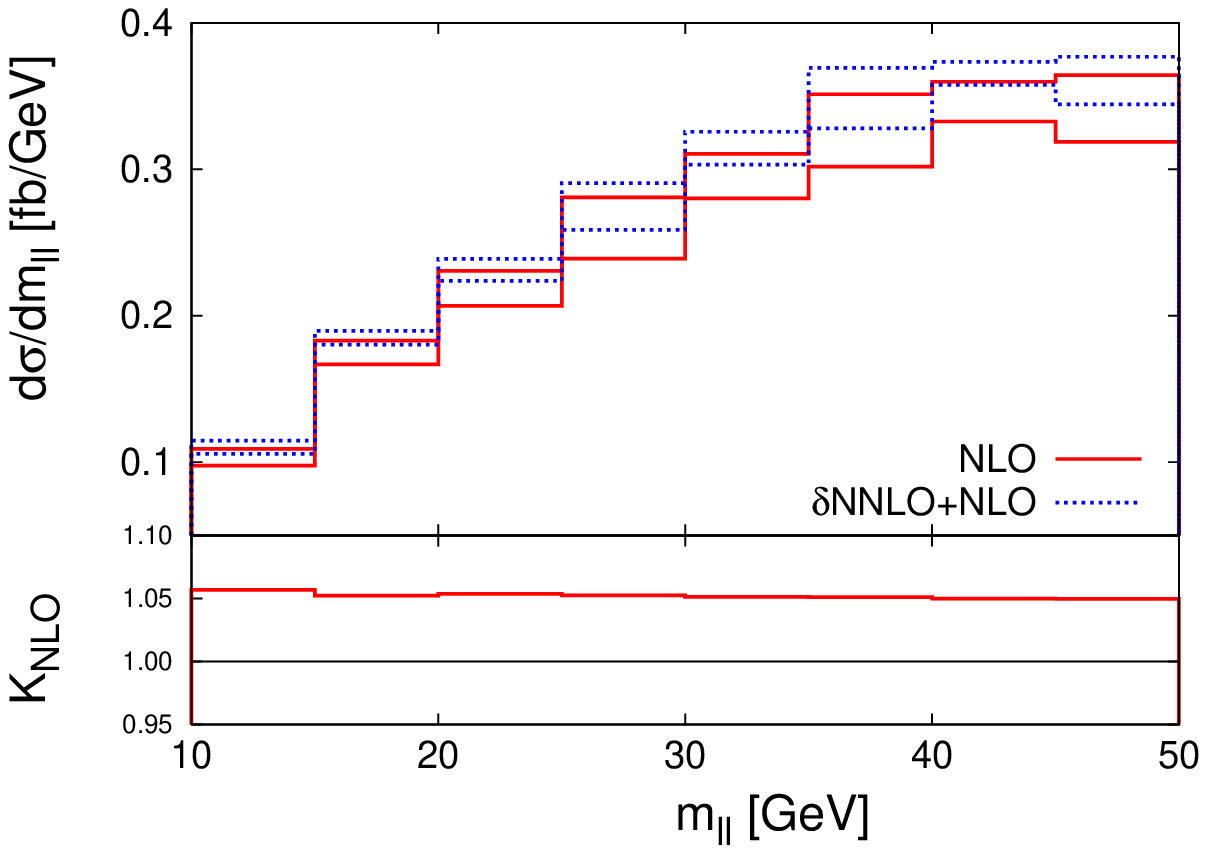}\hspace*{0.1cm}
\includegraphics[scale=0.6]{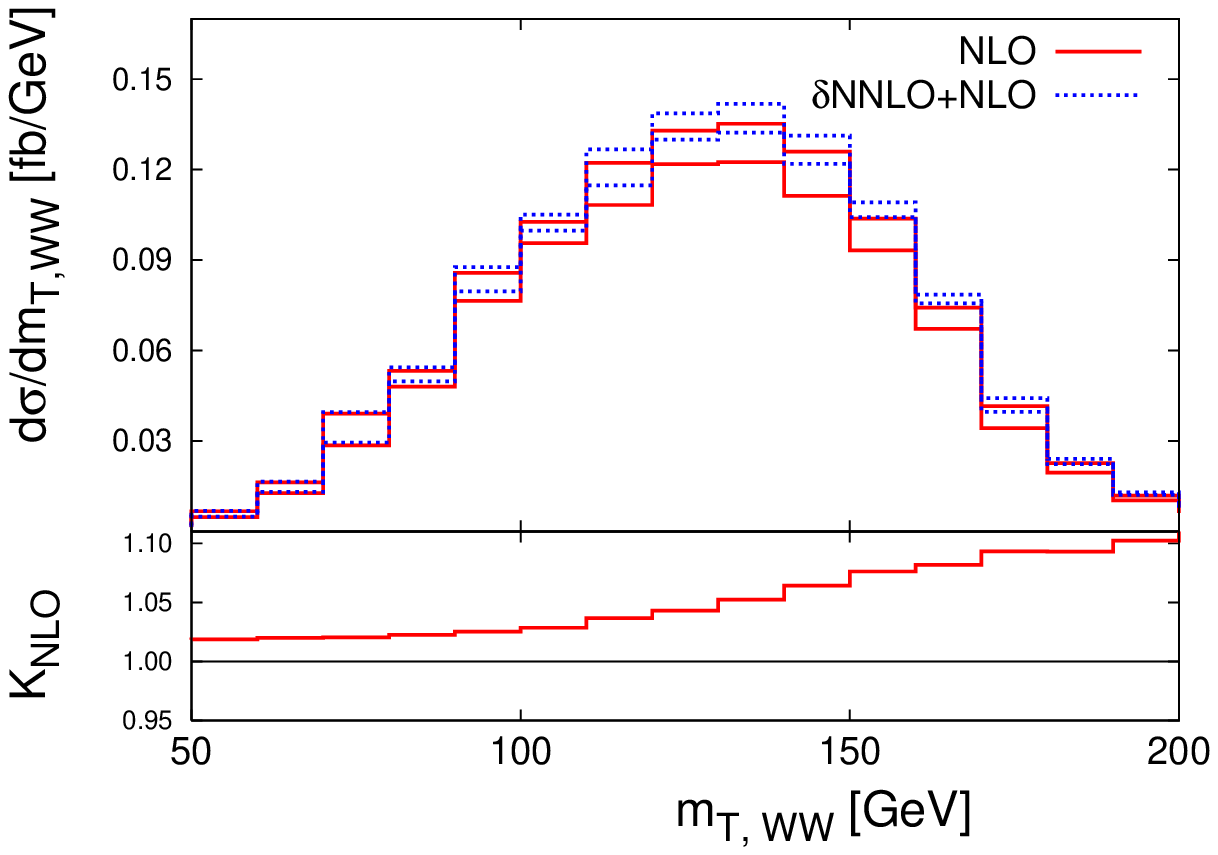}\\
\includegraphics[scale=0.6]{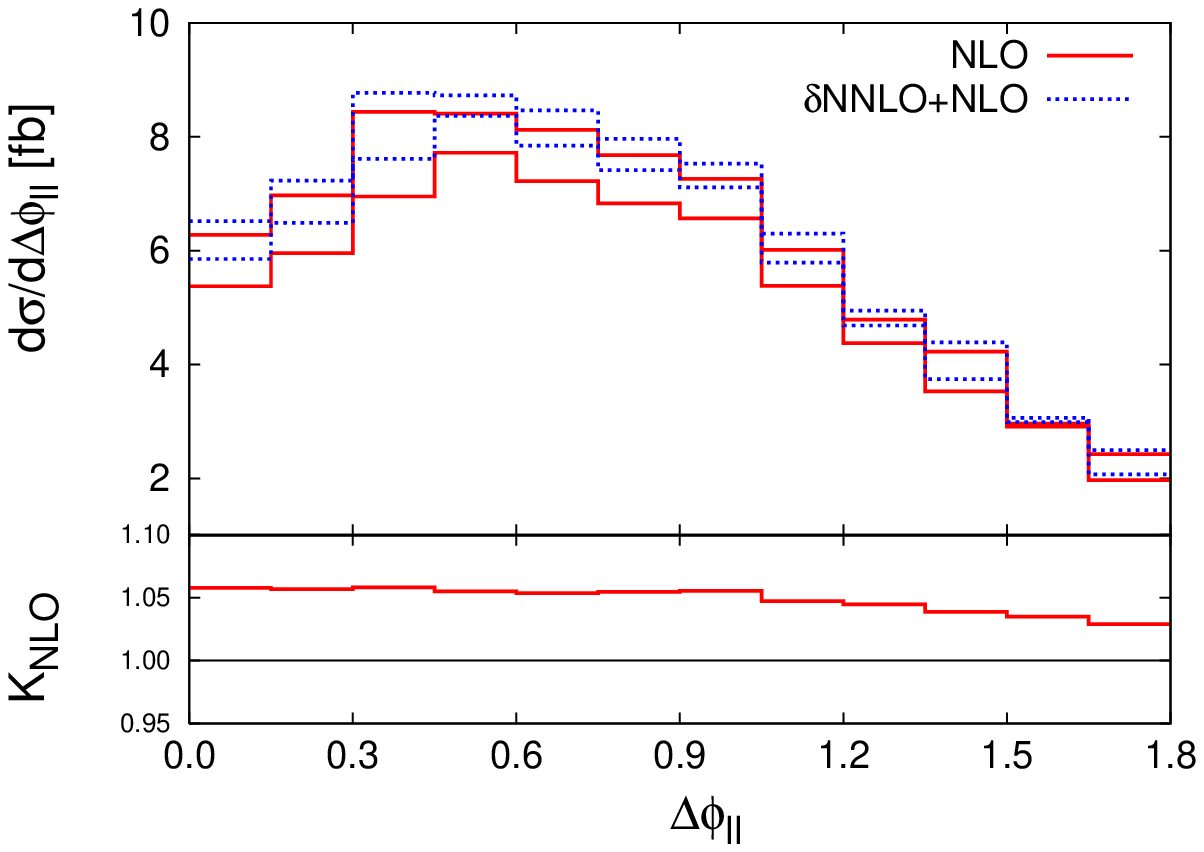}\hspace*{0.1cm}
\includegraphics[scale=0.6]{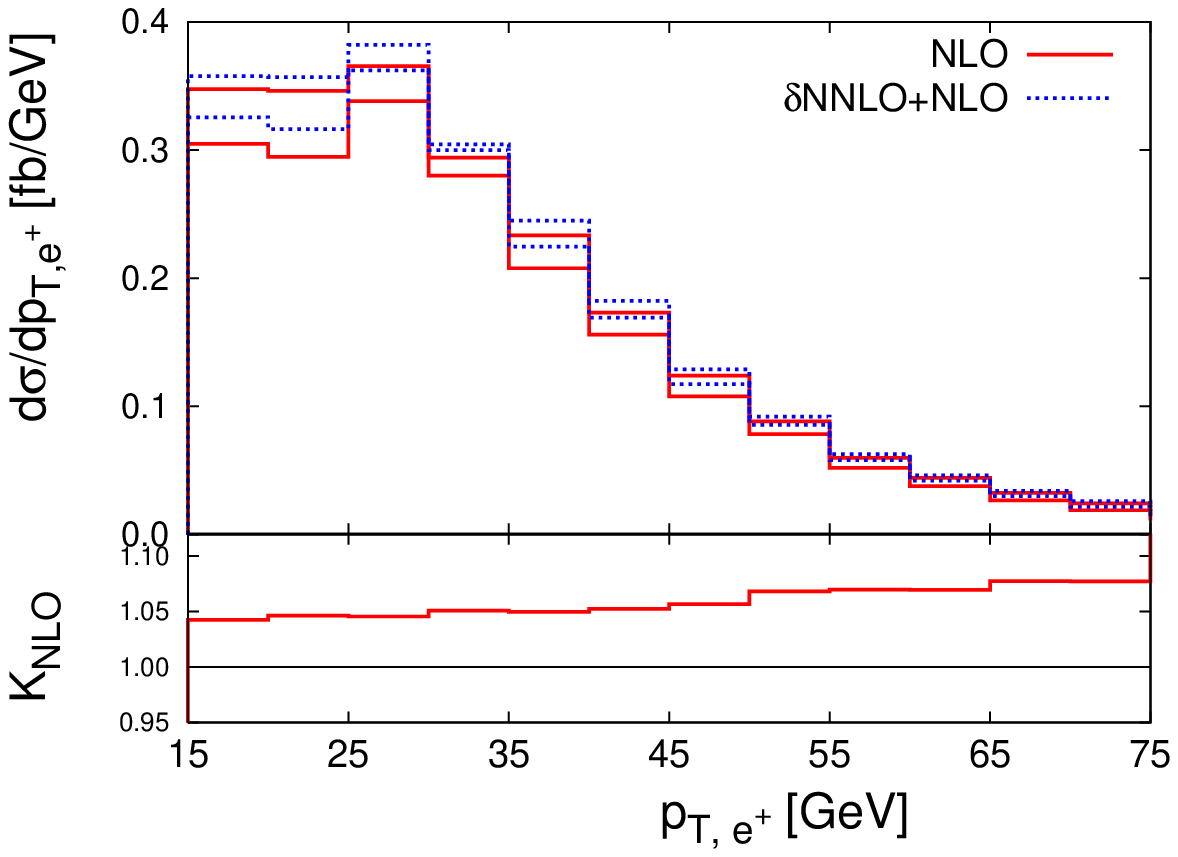}\\
\caption{Distributions of the mass of the charged lepton system
  $m_{ll}$, transverse mass of the $W$-pair $m_{T,WW}$, the azimuthal
  angle between the leptons $\Delta \phi_{ll}$ and the transverse
  momentum of the positron $p_{T, e^+}$, shown at NLO accuracy
  with and without the fermion loop NNLO contribution. We use the Higgs
  search cuts as described in the text, and we display results for
 the $8$ TeV LHC. The upper and lower
  bands show the maximum and minimum deviations from the central scale
  value $2m_W$. 
The ratio $K_{\rm NLO}$ as defined in 
Eq.(3.3)
  is also
  displayed.
}
\label{fig:distr}
\end{center}
\end{figure}

We note that the gluon fusion corrections to $pp \to W^+W^-$ that we
find in this paper are significantly smaller than ${\cal O}(30\%)$
effects reported in Refs.~\cite{Binoth:2005ua,Binoth:2006mf}.  This
happens because the cuts used in those references are more aggressive
than the cuts  that we use here. In particular, stronger cuts on $\Delta
\phi_{ll}$, $m_{ll}$ and, especially, on the transverse momenta of the
charged leptons, account for the larger $gg \to WW$ contribution.  We
have checked that with cuts similar to those adopted in
Ref.~\cite{Binoth:2005ua,Binoth:2006mf} we indeed would have a
significant increase in the relative importance of the gluon fusion
contribution.

\begin{table}[t] 
\begin{center}

Higgs search cuts 

\vspace{0.3cm}

\begin{tabular}{|c|c|c|c|c|c|c|}
  \hline
  & &$\sigma_{\mathrm{LO}}$ & $\sigma_{\mathrm{NLO}}^{\mathrm{incl}}$&
  $\sigma_{\mathrm{NLO}}^{\mathrm{excl}}$ & $\delta
  \sigma_{\mathrm{NNLO}}$ & $ \delta \sigma_{\mathrm{NNLO}} /
  \sigma_{\mathrm{NLO}}^{\mathrm{excl}}$ \\
  \hline
  \multirow{2}{*}{8 TeV}  & $WW$  & $35.6(1)^{+0.9}_{-1.3} $     &
  $51.1(1)^{-0.4}_{+0.9}$      & $38.8(1)^{+1.0}_{-0.8}$  &  $2.7(1)
  ^{-0.5}_{+0.7}$ &  $
7.0\%$ \\
  & $WW$j  & $12.6(1)^{-1.5}_{+1.8}$      &  $10.8(1)^{+0.3}_{-0.7}$
  & $10.6(1)^{+0.3}_{-0.9}$ &  $0.6(1)_{+0.2}^{-0.2}$ & $
5.7\%$\\ 
  \hline
  \multirow{2}{*}{14 TeV} & $WW$  & $63.4(1)^{+3.9}_{-4.7}$      &
  $91.9(2)^{-0.1}_{+0.4}$      & $63.4(2)^{+2.1}_{-2.0}$  &
  $7.5(1)_{+1.5}^{-1.2}$ & $
11.8\%$\\
  & $WW$j  & $28.7(1)^{-2.6}_{+2.9}$      &  $21.6(1)^{+1.2}_{-2.1}$
  & $20.5(1)^{+1.7}_{-2.2}$ & $1.8(2)^{-0.5}_{+0.7}$ & $  
8.8\%$\\
\hline
\end{tabular}
\end{center}
\caption{As for table 1, but using the ATLAS Higgs search cuts as
  described in the text. The ratio in the last column is taken using
  the \textit{exclusive} NLO cross-section.}
\label{tab:xsatlas}
\end{table}

Finally, the integrated cross-section -- even when computed with cuts
optimized for the Higgs boson searches -- is not the whole story since
experimentalists rely on shapes of distributions to distinguish signal
and backgrounds.  In Fig.~\ref{fig:distr} we show the relevant
kinematic distributions for the background $W^+W^-+j$ process, paying
particular attention to the modifications caused by the gluon fusion
contribution.  We display distributions in the azimuthal angle between
the charged leptons, the invariant mass of the charged leptons, the
transverse momentum of the positron, and the transverse mass of the
$W^+W^-$ pair defined as
\begin{equation}
  m_{T,WW} = \sqrt{ (E_{T,ll} + p_{T,\mathrm{miss}})^2 - | {\bf p}_{T,\mathrm{miss}} +  {\bf p}_{T,ll} |^2}.
\end{equation}
Here $E_{T,ll}=\sqrt{ |{\bf p}_{T,ll}|^2 + m_{ll}^2 }$ and
$p_{T,\mathrm{miss}} = | {\bf p}_{T,\mathrm{miss}} |$.  Since spin
correlations in $H \to W^+W^-$ decays force the two charged leptons to be
close to each other, the most interesting, signal-rich regions of the
$\Delta \phi_{ll}$-  and $m_{ll}$-distributions are at
small values of the corresponding variables. The distribution in the
transverse mass of the $W^+W^-$ pair is also useful because
restricting this variable to values comparable to the Higgs boson mass
removes the interference between the Higgs signal and the background
from direct $W^+W^-$ production \cite{Campbell:2011cu}. Furthermore,
ATLAS collaboration uses the distribution of $m_{T,WW}$
to test for the
presence of a signal \cite{ATLAS:2012}.   
In the lower panes in Fig.~\ref{fig:distr},
we show the differential $K$-factors defined as
\begin{equation}
  K_{\rm NLO}  = 
\frac{ {\rm d} \sigma_{\rm NLO+\delta NNLO}}{ {\rm d} \sigma_{\rm NLO} }
\Bigg |_{\mu = 2 m_W}.
\label{eqknlo}
\end{equation}
We notice that the $K_{\rm NLO}$ factor spread over the distributions
$m_{ll}$, $\Delta \phi_{ll}$ and $p_{T, e^+}$ is close to uniform,
while the relative importance of the $\delta \sigma_{\mathrm{NNLO}}$
contribution increases with $m_{T,WW}$. 


\section{Conclusion} \label{sec:concl}

In this paper, we considered the gluon fusion contribution to the
production of a pair of $W$-bosons and a jet at the LHC. While
formally part of the NNLO QCD correction to the $pp \to W^+W^-+$jet
process, this contribution can be treated separately because it is
finite and gauge invariant.  In addition, gluon fusion contributions are
enhanced by the large gluon flux at the LHC and may therefore be  an
important part of the backgrounds to Higgs boson searches.  
In fact, it was found
in Refs.~\cite{Binoth:2005ua,Binoth:2006mf} that, for a particular set
of cuts, the $gg \to W^+W^-$ contribution becomes the largest
radiative correction to $pp \to W^+W^-$, changing the leading order
contribution by as much as $30\%$.

In the current paper, we used the set of cuts employed by the ATLAS
collaboration in their current searches for the Higgs boson to
estimate the gluon fusion contribution to $pp \to W^+W^-+0~{\rm
  or}~1~$jet and we find significantly smaller radiative corrections
than the results quoted in Refs.~\cite{Binoth:2005ua,Binoth:2006mf}.
This is a combined effect of looser cuts on $\Delta \phi_{ll}$,
$m_{ll}$ and the lepton transverse momenta currently used by the
ATLAS collaboration compared to those used for the discussion in
Refs.~\cite{Binoth:2005ua,Binoth:2006mf}.  For the ATLAS Higgs search
cuts we find that the gluon fusion contribution to $pp \to W^+W^-j$ is somewhat
larger than, but still comparable to, the scale uncertainty of the
exclusive NLO cross-section and most of the kinematic
distributions. However, as our results show, these conclusions very
much depend on the exact cuts applied.  Should the experimental cuts
change significantly, the calculations reported in this paper should
be reconsidered.

{\bf Acknowledgments} K.M. acknowledges useful conversations with
F. Petriello about the importance of gluon fusion background processes
for the Higgs boson searches.  This research is supported in part by
NSF grants PHY-0855365, by the DOE grant DE-AC02-06CD11357, by the
LHCPhenoNet network under the Grant Agreement PITN-GA-2010-264564, and
by the British Science and Technology Facilities Council.  M.S. is
grateful for support from the Director's Fellowship of Argonne
National Laboratory.

\appendix 

\section{Appendix}
In this appendix, we display the results for some of the primitive
amplitudes, as well as a full amplitude summed over helicities, for a
single phase space point. We consider the process $gg \to W^+(\to
\nu_e e^+) W^-(\to \mu^- \bar{\nu}_{\mu}) g$, and use the phase-space
point $(E,p_x,p_y,p_z)$ (in GeV):
\begin{small}
\begin{equation}
 \begin{split}
  p_{g_1} &=  (-500, 0,0,-500),\\
  p_{g_2} &=  (-500,0, 0,500),\\
  p_{g_3} &=   (86.3540681437814,-15.2133893202618,37.6335512949163,-76.2187226821854),\\
  p_{\nu_e} &=  (414.130068374543,232.145564945939,332.7544367808,-82.9857518524426),     \\                                        
  p_{e^+} &=  (91.8751521026384,-43.3570226791011,-24.0058236140057,77.3623460793435),    \\  
  p_{\mu^-} &=  (280.118181809376,-83.1261116505822,-263.203856758651,47.7490851160266),   \\                                         
  p_{\bar{\nu}_{\mu}} &=  (127.522529569661,-90.4490412959935,-83.1783077030789,34.0930433392580).
 \end{split}
\end{equation}
\end{small}
We present the results for the primitive amplitude $A_f(g_1,g_2,g_3)$
as defined in Eq.~\eqref{partamp} for four sets of helicities of the
three gluons. In Table \ref{tab:app1}, we display both massless and
massive fermion loop amplitudes as defined in Eq.~\eqref{0m}. We show
only the finite parts of the amplitude, since there are no poles.
\begin{table}
 \begin{center}
 \begin{tabular}{|c|c|c|}
 \hline
& $|A_{f,0}|$ & $|A_{f,m}|$ \\
\hline
$g_1^+,g_2^+, g_3^+$ & $3.844351 $ & $0.9542655$\\
$g_1^+,g_2^+, g_3^-$ & $5.699256$ & $1.606166$\\
$g_1^+,g_2^-, g_3^-$ & $3.522647$ & $6.204687$\\
$g_1^-,g_2^-, g_3^-$ & $ 23.21330 $ & $26.65234$\\
\hline
\end{tabular}
\end{center}
\caption{
  Numerical results for the massless and massive fermion loop primitive
  amplitudes $A_{f,0}(g_1,g_2,g_3)$ and $A_{f,m}(g_1,g_2,g_3)$, in units
  of $10^{-6} \mathrm{GeV}^{-3}$.} 
\label{tab:app1}
\end{table}

We also give results for the full amplitude squared, that is, the
color-squared amplitude from Eq.~\eqref{partamp} summed over all gluon
helicities. If we consider only one massless generation in the loop,
the result is $8.348897 \times 10^{-8} \mathrm{GeV}^{-6} $. If we
consider one massive generation in the loop, the result is $1.128414
\times 10^{-7} \mathrm{GeV}^{-6}$. For the setup used in deriving our
results, i.e. two massless and one massive generation in the loop, the
amplitude squared is $7.830968 \times 10^{-7} \mathrm{GeV}^{-6}$.

\end{document}